# 薄分子云湍流结构函数的倾角改正及其应用*

钱 磊[1,2†]

(1 中国科学院国家天文台 北京 100101)
(2 中国科学院大学天文与空间科学学院 北京 100049)

**摘要** 通过结构函数可以测量湍流的能量级联速率. 在实际观测中, 无法测量分子云中气体的3维速度, 这使得其湍流结构函数难以测量. 对垂直于视线方向的薄分子云的情形, 结构函数$S_{tt}^2$可以通过云核速度弥散(core velocity dispersion, CVD)进行测量, $\text{CVD}^2 = \frac{1}{2}S_{tt}^2$. 对此进行推广, 对于不垂直于视线方向的薄分子云, $\text{CVD}^2 = \frac{1}{2}S_{tt}^2\left(1 - \frac{1}{8}\cos^2\theta\right)R^{2/3}$, 其中, $\theta$是视线方向与投影方向的夹角, 平均投影距离与3维距离之比$R$可以用第2类椭圆积分$E(k,\varphi)$表示为$R = \frac{2}{\pi}E(\cos\theta, \frac{\pi}{2})$.

**关键词** 星际介质: 分子云, 星际介质: 运动学和动力学, 星际介质: 谱线
**中图分类号:** P141; **文献标识码:** A

## 1 引言

湍流是分子云演化中重要的物理过程[1–2]. 一方面, 湍流密度涨落在分子云中形成了高密度区域, 成为云核的种子. 另一方面, 湍流在云核中抵抗了引力塌缩, 阻碍了恒星的形成. 在湍流较强的区域, 恒星形成效率显著降低[3].

分子云中湍流的能量有可能是通过星系盘的较差转动[4]、星系盘的潮汐作用[5]、星系盘中的大尺度引力不稳定性[6–7]、超新星爆发[8–10]等机制从大尺度注入. 随后能量从大尺度传递到小尺度, 这个过程被称为能量级联[11]. 级联传递的能量最终在小尺度耗散. 对于不可压缩湍流, 能量级联速率和能量耗散率相等, 测定了能量级联速率, 就知道了能量耗散率. 而对于可压缩湍流, 能量级联速率也能对能量耗散率给出限制.

由于观测条件限制, 我们通常只能测量分子云中气体运动的视向速度和分子云的横向尺度. 分子云中湍流能量耗散率难以直接测量, 通常使用半解析公式结合数值模拟进行估计. 对于垂直于视线方向薄分子云(厚度小于横向尺度1/10的分子云[12])的特殊情形, Qian等[13]提出了一种用观测量得到结构函数, 直接测量湍流能量级联速率, 从而得出湍流能量耗散率的方法, 并用这种方法测量了金牛座分子云中的湍流能量耗散率

---

为$(0.45 \pm 0.05) \times 10^{33}$ erg·s$^{-1}$. 在这里, 分子云的横向尺度可以通过观测到的分子云角尺度和距离得到. 而分子云的厚度可以通过观测分子云中的气泡结构的形态[14]、分子云中的线宽-尺度关系[12]等方法进行估计. 这种用观测量得到结构函数的方法只适用于薄分子云. 在厚的分子云中, 观测的云核速度弥散不随横向尺度变化[12], 无法得到结构函数.

实际的分子云通常和视线方向有夹角, 这个夹角可以通过测量分子云中不同部分的年轻恒星的距离大致估计[15]. 故而需要将结构函数的测量推广到薄分子云和视线方向不垂直的情形. 在此情形, 云核速度弥散与结构函数$S_{tt}^2$之比有一个依赖于角度的改正因子, 这个改正因子会影响湍流能量耗散率的测量. 接下来推导这个改正因子.

## 2 结构函数及相关研究更新

### 2.1 纵向结构函数和横向结构函数

薄分子云的几何如图1所示. $\vec{s}$是视线方向. $\vec{l}$是连接点1和点2的矢量, $\vec{n}$是平面的法线方向, $\vec{t}$垂直于$\vec{l}$. 2阶纵向结构函数$S_{ll}^2$和2阶横向结构函数$S_{tt}^2$分别定义为:

$$S_{ll}^2(l_{12}) = \langle (v_{l2} - v_{l1})^2 \rangle = \langle \delta v_l^2 \rangle, \tag{1}$$

和

$$S_{tt}^2(l_{12}) = \langle (v_{t2} - v_{t1})^2 + (v_{n2} - v_{n1})^2 \rangle = 2\langle \delta v_t^2 \rangle, \tag{2}$$

其中$v$是速度, $\delta v$是速度差, 下标1、2表示点1和点2的物理量. 下标$l$、$t$、$n$分别表示物理量在$\vec{l}$、$\vec{t}$、$\vec{n}$方向的分量(图1). $l_{12}$是连接点1和点2的线段. 为了简洁, 后文$l_{12}$记作$l$. 纵向和横向结构函数的关系为[16]:

$$S_{tt}^2 = 2\left(1 + \frac{l}{2}\frac{\partial}{\partial l}\right) S_{ll}^2. \tag{3}$$

在不可压缩湍流的惯性区, 单位质量的能量级联速率$\epsilon$与$S_{ll}^2$和$S_{tt}^2$有确定的关系[17]:

$$S_{ll}^2 = C_{12}\epsilon^{2/3}l^{2/3}, \tag{4}$$

和

$$S_{tt}^2 = \frac{8}{3}C_{12}\epsilon^{2/3}l^{2/3}, \tag{5}$$

其中$C_{12} = 2.12$是一个普适常数.

### 2.2 与视线方向有一定夹角的薄云的结构函数

假定一块薄分子云和视线方向的夹角为$\theta$, 云中两点连线矢量记作$\vec{l}$, 云平面内垂直于$\vec{l}$的方向记作$\vec{t}$, 云平面的法线方向记作$\vec{n}$, 视线方向记作$\vec{s}$. $\vec{l}$和视线方向在云平面投影方向的夹角(即方位角)记作$\varphi$. 则两点的速度差在视线方向的分量为

$$\delta v_s = \delta v_l \cos\varphi \cos\theta + \delta v_t \sin\varphi \cos\theta + \delta v_n \sin\theta, \tag{6}$$

方程两边平方后取平均, 由于$\langle \delta v_l \delta v_t \rangle$等交叉项的平均值为0, 可以得到

$$\langle \delta v_s^2 \rangle = \langle \delta v_l^2 \rangle \overline{\cos^2 \varphi \cos^2 \theta} + \langle \delta v_t^2 \rangle \overline{\sin^2 \varphi \cos^2 \theta} + \langle \delta v_n^2 \rangle \overline{\sin^2 \theta}$$

$$= \frac{S_{ll}^2}{2} \cos^2 \theta + \frac{S_{tt}^2}{2} \left( \frac{\cos^2 \theta}{2} + \sin^2 \theta \right)$$

$$= S_{tt}^2 \left( \frac{1}{2} - \frac{1}{16} \cos^2 \theta \right) ,$$

其中带上横线的项表示对角度$\varphi$平均, $\langle \delta v_t^2 \rangle = \langle \delta v_n^2 \rangle = \frac{S_{tt}^2}{2}$, $\langle \delta v_l^2 \rangle = S_{ll}^2 = \frac{3}{8} S_{tt}^2$. 云核速度弥散定义为CVD$\equiv \sqrt{\langle \delta v_s^2 \rangle}$, 其平方就是方程左边的项.

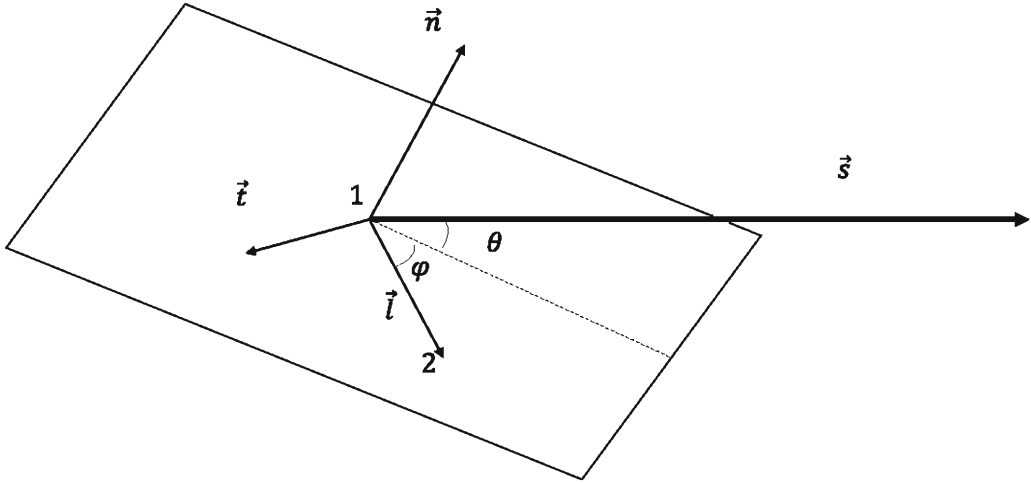

图 1  薄分子云的几何示意图. $\vec{s}$是视线方向. $\vec{l}$是连接点1和点2的矢量, $\vec{n}$是平面的法线方向, $\vec{t}$垂直于$\vec{l}$. 视线方向和分子云平面的夹角为$\theta$. $\varphi$为方位角.

Fig. 1  The sketch of a thin molecular cloud. $\vec{s}$ is the line of sight. $\vec{l}$ is the vector connecting point 1 and 2. $\vec{n}$ is the normal vector of the plane. $\vec{t}$ is perpendicular to $\vec{l}$. The intersection angle of the molecular cloud and the line of sight is $\theta$. $\varphi$ is the azimuth angle.

### 2.3  投影距离

注意到前面给出的公式中, 两点间距离$l$是3维距离. 实际观测中, 我们能看到的距离是两点在垂直于视线方向的面, 即天球面上的投影距离$L$. 简单的计算可以得到:

$$L = \sqrt{l^2 \sin^2 \varphi + l^2 \cos^2 \varphi \sin^2 \theta}$$
$$= l \sqrt{1 - \cos^2 \theta \cos^2 \varphi} .$$

定义平均投影距离与3维距离之比$R \equiv \frac{\overline{L}}{l}$, 其中$\overline{L}$是$L$对角度$\varphi$的平均, 计算得到:

$$R = \frac{\int_0^{2\pi} \sqrt{1 - \cos^2 \varphi \cos^2 \theta} \, \mathrm{d}\varphi}{2\pi}$$
$$= \frac{2}{\pi} E \left( \cos \theta, \frac{\pi}{2} \right) .$$

其中$E(k,\varphi) \equiv \int_0^\varphi \sqrt{1-k^2\sin^2 x}\,\mathrm{d}x$是第2类椭圆积分. $R$随$\theta$的变化见图2, 注意到在视线方向与薄分子云垂直时($\theta=90°$), $R=1$.

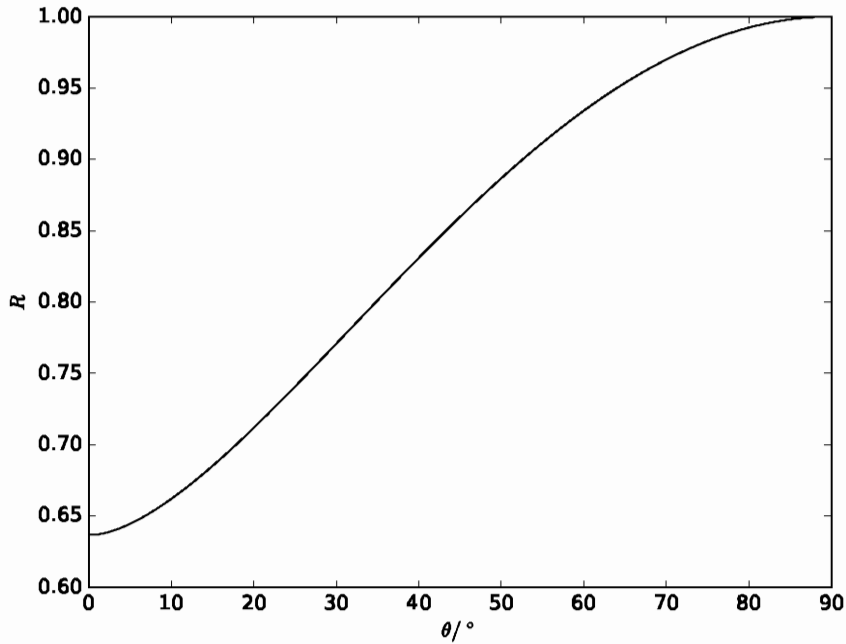

图 2　$R$随视线方向与薄分子云夹角$\theta$的变化. 在视线方向与薄分子云垂直时($\theta=90°$), $R=1$.

Fig. 2　The ratio of the average projected distance and the 3D distance $R \equiv \frac{\overline{r}}{l}$ changes with the intersection angle of the molecular cloud and the line of sight $\theta$. We have $R=1$ when the line of sight is perpendicular to the thin molecular cloud ($\theta=90°$).

### 2.4　结构函数的改正因子

综合结构函数和投影距离的结果, 在薄分子云和视线方向夹角为$\theta$时, 云核速度弥散和结构函数的关系为:

$$\mathrm{CVD}^2 = \frac{1}{2}S_{tt}^2\left(1-\frac{1}{8}\cos^2\theta\right)R^{2/3}. \tag{7}$$

$\left(1-\frac{1}{8}\cos^2\theta\right)R^{2/3}$为改正因子. 在薄分子云和视线方向垂直时, 此改正因子为1, 回到文献[13]中不考虑厚度改正的结果.

### 2.5　金牛座分子云的湍流耗散率

金牛座分子云相对视线方向的方位是不确定的, 有一些迹象表明金牛座分子云与视线方向不垂直[18]. 根据上面得到的结构函数的改正因子可以得到能量耗散率的改正因子$\left(1-\frac{1}{8}\cos^2\theta\right)^{-3/2}R^{-1}$, 结合金牛座湍流耗散率已有工作[13]可以得到改正后的能量耗散率.

由于不知道金牛座分子云与视线方向的准确夹角$\theta$, 我们对几个夹角$\theta=30°$、$45°$、$60°$分别进行了计算, 得到的湍流耗散率分别为$(0.68\pm0.08)\times10^{33}$ erg·s$^{-1}$、$(0.58\pm0.06)\times10^{33}$ erg·s$^{-1}$、$(0.51\pm0.06)\times10^{33}$ erg·s$^{-1}$.

## 3　总结与展望

分子云湍流结构函数的测量通常需要测量速度的3维分量. 在特殊情形下, 可以通过视向速度的测量给出结构函数. 分子云的厚度是一个关键因素, 研究表明对于厚度较大的分子云, 无法通过视向速度测量得到结构函数[12].

对于薄分子云, 当分子云和视线方向垂直时, 云核速度弥散和结构函数$S_{tt}^2$有简单的正比关系. 在薄分子云和视线方向不垂直时, 存在一个依赖角度$\theta$的改正因子$\left(1-\frac{1}{8}\cos^2\theta\right)R^{2/3}$, 对应的能量耗散率的改正因子为$\left(1-\frac{1}{8}\cos^2\theta\right)^{-3/2}R^{-1}$.

未来测量一般分子云中湍流的结构函数, 需要测量分子云的3维结构以及3维速度. 测量分子云中气体的3维运动的一个可能性是用分子云中的年轻恒星进行测量. 目前通过Gaia卫星发布的星表已经可以得到大量年轻恒星的3维位置和3维速度. 年轻恒星的运动可能部分反映气体的运动. 另一方面, 年轻恒星可以作为背景光源, 通过吸收线研究分子云中的气体. 未来的工作有待更深入研究目前已有的吸收谱观测以及詹姆斯·韦伯太空望远镜(James Webb Space Telescope, JWST)所能起到的作用.

# The Inclination Correction to the Turbulence Structure Function of Thin Molecular Clouds and Its Application

QIAN Lei[1,2]

(1 National Astronomical Observatories, Chinese Academy of Sciences, Beijing 100101)
(2 School of Astronomy and Space Science, University of Chinese Academy of Sciences, Beijing 100049)

**ABSTRACT**　The energy cascade rate of turbulence can be measured with the structure function. In practice, the 3D velocity of the gas in molecular cloud is hard to measure, which makes the measurement of structure function difficult. In the case of thin molecular clouds perpendicular to the line of sight, the structure function $S_{tt}^2$ can be measured with core velocity dispersion (CVD), $\mathrm{CVD}^2 = \frac{1}{2} S_{tt}^2$. This method was extended to the case when the thin molecular cloud is not perpendicular to the line of sight, with intersection angle $\theta$, $\mathrm{CVD}^2 = \frac{1}{2} S_{tt}^2 \left(1 - \frac{1}{8}\cos^2\theta\right) R^{2/3}$, where $R$ is the ratio of the average projected distance to the 3D distance, can be expressed with elliptic integrals of the second kind $E(k, \varphi)$ as $R = \frac{2}{\pi} E(\cos\theta, \frac{\pi}{2})$.

**Key words**　ISM: clouds, ISM: kinematics and dynamics, ISM: lines and bands